# Pulse thermal imaging of FUHAO bronze artifact

Li Wang, Ning Tao, Wei Liu, Xiaoli Li, Yi He, Xue Yang, Jiangang Sun, Cunlin Zhang

**Abstract**: The accurate identification of historical restoration traces and material degradation is essential for the scientific preservation of ancient bronzes. In this study, the prestigious FUHAO bronze artifact (late Shang period, 13[th]-11[th] century BCE) was non-destructively examined using pulsed thermal imaging (PT). By combining single- and double-layer heat conduction models with Thermal Tomography (TT), this approach allowed for precise spatial localization of repair crevices, patches, and filler materials, while also distinguishing restorative interventions from the original bronze substrate. The artifact was revealed to have been assembled from multiple fragments, exhibiting uneven surface corrosion and clear evidence of prior conservation. The results not only provide direct insights for conservation strategy and historical interpretation but also demonstrate the capability of pulsed thermal imaging as an effective diagnostic tool for the integrated surface and subsurface assessment of cultural heritage objects.

**Keywords:** Pulsed thermal imaging, Bronze artifact, Cultural heritage diagnostics, Non-destructive evaluation, Thermal tomography

## 1 INTRODUCTION

Infrared thermography (IRT) is a non-destructive testing (NDT) technique based on the evaluation of surface temperature evolution, used to detect surface and subsurface material inhomogeneities without contact. Depending on whether external excitation is applied, IRT is classified as either passive or active. Active thermography, which introduces controlled thermal stimulation, has seen extensive development and includes several modalities. Pulsed thermography (PT) [1] applies a short, high-energy thermal pulse to the sample surface, and the subsequent temperature decay is analyzed to probe subsurface features. Lock-in thermography (LIT) [2] uses periodic thermal waves and extracts defect information from the amplitude and phase lag of the thermal response. Step-heating thermography (ST) [3–4] involves continuous heating while monitoring the surface temperature rise over time. These active approaches [5–7] are widely applied for the qualitative and quantitative identification of defects and have gained increasing importance in the field of cultural heritage conservation, where recent studies have demonstrated their successful application in detecting delamination in paintings and cracks in mural substrates [13–15]. For metallic artifacts, however, a technique with both high temporal resolution and sensitivity to subtle thermal property variations is required [16].

Given the irreplaceable value of cultural artifacts, non-destructive diagnostic methods are essential for their preservation. Although X-ray imaging [8–10] is routinely used to visualize internal structures, it faces limitations when examining objects with complex geometries or overlapping material layers, which can obscure critical features. Other imaging techniques, such as multispectral imaging (MSI) [11] and hyperspectral imaging (HSI) [12], offer surface characterization but provide limited depth penetration. Safety concerns and practical constraints further restrict the applicability of some methods in heritage conservation contexts. These challenges highlight the need for alternative NDT approaches that are safe, depth-sensitive, and capable of mapping material variations in complex artifacts.

Pulsed thermography (PT) presents a promising alternative in this regard. By delivering a short, intense flash to thermally excite the surface and recording the resulting temperature evolution with an infrared camera, PT enables the detection of internal interfaces, defects, and material inhomogeneities based on their thermal properties. To extend its analytical capability, thermal tomography (TT) can be applied to reconstruct depth-resolved effusivity profiles, offering a three-dimensional perspective on subsurface structure.

In this study, we apply pulsed thermal imaging combined with thermal tomography and one-dimensional heat transfer modeling to investigate the prestigious "Fu Hao" bronze artifact from the late Shang period. The objectives are to: (1) identify and spatially locate previous restoration interventions such as repair patches and crevices; (2) differentiate restorative materials from the original bronze substrate; and (3) characterize the distribution and morphology of surface corrosion. Through this integrated approach, we demonstrate how PT and TT can serve as effective diagnostic tools for the non-destructive evaluation of bronze cultural heritage, providing insights essential for conservation planning and historical analysis.

## 2 ARTIFACT AND METHODS

### 2.1. "Fu Hao" bronze artifact

The "Fu Hao" bronze double artifact (late Shang dynasty, c. 13th–11th century BCE), housed in the National Museum of China, is a major ceremonial vessel excavated from the Fu Hao tomb at Yinxu (Anyang, Henan) in 1976. It comprises two principal components: a cap and a body. The body has a roughly rectangular opening, each long side decorated with a prominent animal head (Fig. 1), while each short side features an elephant head (Fig. 2). The surface is adorned with intricate patterns, including birds and gluttony motifs. The mid-ridge of the lid and its four sloping corners are ribbed, and the central protrusion bears an owl-face design that aligns with the notch on the body. The overall dimensions are substantial: height ~610 mm, length ~890 mm, and outer width ~260 mm. The cap is shaped like a house roof.

Figures 1 and 2 reveal a heterogeneous surface covered with various forms of corrosion (patina). The complexity of the surface is increased by the uneven thickness of the corrosion layer and the presence of obscured decorative details. When excavated, the artifact was found adjacent to a tomb wall, resulting in asymmetric corrosion: the side facing the wall (termed the "back") exhibits more severe corrosion than the opposite ("front") side.

Notably, the left elephant head is a modern replica, as the original was lost. Unlike the authentic right elephant head, which shows decorative patterns behind the ear, the replica lacks such detailing—a difference discernible by visual inspection. However, most restoration traces, such as repair crevices and patches, are not visually obvious.

Historical photographs indicate that the artifact underwent multiple restoration campaigns after its discovery and was originally found fragmented and damaged [18, 19]. Identifying these interventions is critical for assessing structural stability and planning conservation, yet visual inspection alone is insufficient. This underscores the need for advanced non-destructive techniques such as the pulsed thermography approach presented here.

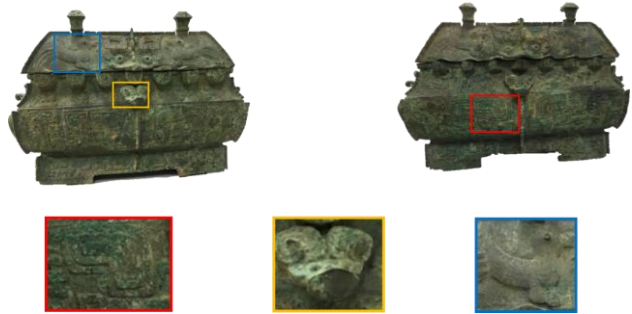

Fig. 1 Artifact Surfaces: Front, Back, and Detailed Views.

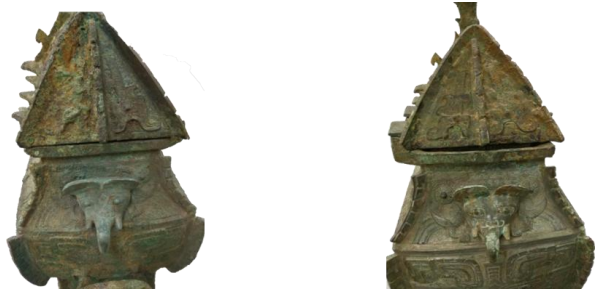

Fig. 2 The appearance characteristics of sides of the bronze artifact.

Historical photographs indicate that the artifact underwent multiple restoration campaigns after its discovery and was originally found fragmented and damaged [18, 19]. Identifying these interventions is critical for assessing structural stability and planning conservation, yet visual inspection alone is insufficient. This underscores the need for advanced non-destructive techniques such as the pulsed thermography approach presented here.

**2.2. Pulse thermal imaging**

   **2.2.1 experimental set-up and data acquisition**

The experimental setup, shown schematically in Figure 3, was configured as follows. Two high-energy flash lamps (3) served as the thermal excitation source. Each lamp had a pulse width of 0.001 s and a maximum energy output of 4.8 kJ, and they were fixed symmetrically on the inner panel of a light hood to illuminate the sample uniformly. The thermal response was captured by a FLIR infrared camera (2) with a resolution of 320 × 256 pixels and a spectral response range of 7.7–9.3 μm. An optical filter (1) was placed in front of the camera lens. The object under test—the bronze artifact (4)—was positioned to receive the flash excitation. Following each flash pulse, the camera recorded the surface cooling sequence at a frame rate of 225 Hz, and the data were transferred to a computer for subsequent pulsed thermography (PT) processing.

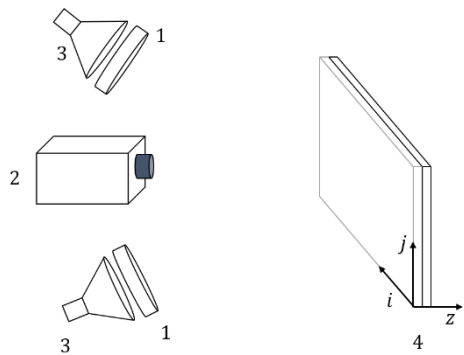

Fig. 3 Schematic of the experimental setup.

   **2.2.2 Single-layer and double-layer material system**

As seen in Figs. 1 and 2, most of the artifact's surface is covered by a corrosion layer, implying that the wall can be modelled as a double-layer system consisting of a corrosion (rust) layer atop a bronze substrate. In regions where the substrate is severely thinned due to corrosion, a three-layer model may be more appropriate. Restoration materials, which have evolved from early copper-alloy repairs to modern epoxy-resin compounds doped with mineral pigments, possess thermal properties distinct from those of bronze. During PT inspection, these repaired patches exhibit different temperature distributions and cooling rates compared to the sound areas. Although the patches may be surface-coloured to match the original patina, the pigment layer is negligible in thickness; therefore, patches can be treated as single-layer structures.

The surface temperature response under PT excitation differs between single-layer and double-layer configurations. The first derivative of temperature with respect to time (or its logarithmic derivative) provides a signature for identifying the layer structure and uniformity of the corrosion layer.

For a single-layer material of finite thickness L subjected to a short heat pulse on its front surface, the surface temperature $T(0, t)$ is given by [20]:

$$T(0,t) = \frac{Q}{e\sqrt{\pi t}}\left[1 + 2\sum_{n=1}^{\infty} \exp\left(-\frac{(nL)^2}{\alpha t}\right)\right] \tag{1}$$

where $Q$ is the absorbed energy density on the surface, $\rho c$ is heat capacity and $\alpha$ is thermal diffusivity, when the $L \to \infty$, the surface temperature $T(0,t)$ correspond to the thermal propagation in a homogeneous semi-infinite solid is:

$$T(0,t) = \frac{Q}{e\sqrt{\pi t}} \tag{2}$$

This implies that after a certain period, the surface temperature of the object becomes constant and depends on its thermal properties. In the case of a single-layer sample, the logarithmic derivative of the surface temperature concerning time, i.e., $dln(T)/dln(t)$,

$$\frac{dlnT}{dlnt} = -\frac{1}{2} + \frac{2\sum_{n=1}^{\infty}\frac{n^2l^2}{\alpha t}\exp\left(-\frac{n^2l^2}{\alpha t}\right)}{1+2\sum_{n=1}^{\infty}\exp\left(-\frac{(nl)^2}{\alpha t}\right)} \tag{1}$$

follows a curve with two distinct stages. Initially, the temperature decays at a rate of -0.5, eventually approaching zero as time becomes large[21].Fig.1 shows a schematic diagram of the variation in the $dlnT/dlnt$ curve for a homogeneous material with an adiabatic back surface after receiving a heat flux on its front surface under ideal conditions. The shape and steepness of this curve depend on the material own thickness $l$ and thermal property $\alpha$.。

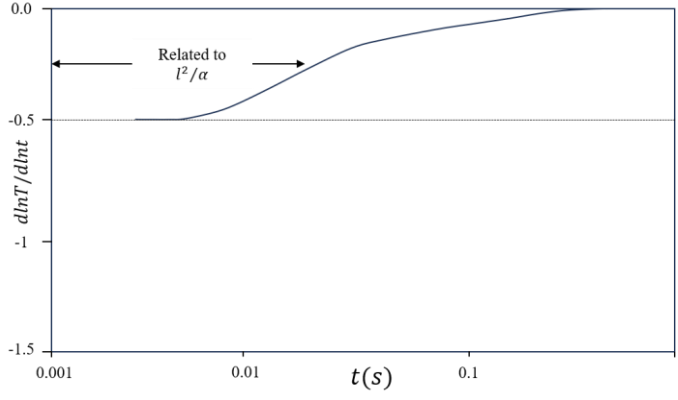

Fig. 4 Characteristics of logarithmic surface of temperature slope data for single-layer material

When considering a sample model comprising of a double-layer structure with a semi-infinite second layer and a first layer with thickness $L_1$, the response of the surface temperature to thermal excitation can be characterized by a specific solution:

$$T_f(t) = \frac{Q}{e_1\sqrt{\pi t}}\left[1 + 2\sum_{n=1}^{\infty}\Gamma^n \exp\left(-\frac{n^2 L_1^2}{\alpha_1 t}\right)\right] \quad (2)$$

where $\Gamma = (e_1/e_2 - 1)/(e_1/e_2 + 1)$, $e$ is the thermal effusivity of material.

In logarithmic form, the characteristics of these solutions can be better observed.

$$\frac{d\ln T}{d\ln t} = -\frac{1}{2} + \frac{2\xi\sum_{n=1}^{\infty} n^2\Gamma^n \exp(-n^2\xi)}{1+2\sum_{n=1}^{\infty}\Gamma^n \exp(-n^2\xi)} = f(e_{12},\xi) \quad (3)$$

Where $f$ is the function of $e_{12} = e_1/e_2$ and $\xi = L_1^2/\alpha_1 t$. When the second layer of material has finite thickness, the situation becomes more complicated, as the reflection from the back surface must be taken into account. Under these conditions, the temperature solution at the front surface of the material is given by[20,23]:

$$T(t) = T_\infty \cdot \left[1 + 2\frac{x_1\omega_1 + x_2\omega_2}{x_1+x_2} \times \sum_{k=1}^{\infty} \frac{x_1\cos(\omega_1\gamma_k) + x_2\cos(\omega_2\gamma_k)}{x_1\omega_1\cos(\omega_1\gamma_k) + x_2\omega_2\cos(\omega_2\gamma_k)} \exp\left(-\frac{\gamma_k^2 t}{\eta_2^2}\right)\right] \quad (4a)$$

$\gamma_k$ is the Kth positive root of the following equation:

$$x_1 \sin(\omega_1\gamma) + x_2 \sin(\omega_2\gamma) = 0 \quad (6b)$$

$x$ and $\omega$ are defined as

$$x_i = e_{12} - (-1)^i, \quad e_i = \sqrt{k_i \rho_i c_i}, \quad i = 1, 2 \quad (6c)$$

$$\omega_i = \eta_{12} - (-1)^i, \quad \eta_i = L_i/\sqrt{\alpha_i}, \quad i = 1,2 \quad (6d)$$

with

$$e_{12} = e_1/e_2, \quad \eta_{12} = \eta_1/\eta_2 \quad (6e)$$

For Equation 5, when its derivative function is equal to 0, it represents a unique $\xi$-value, giving Equation 5 a negative slope[20]. As depicted in Fig.5, the logarithmic curve of the two-layer structure remains constant at -0.5 along the horizontal axis initially, followed by a negative slope whose amplitude correlated to the ratio of $e1/e2$. Upon further analysis of the surface, the location of the negative slope on the horizontal axis can be determined by $L_1^2/\alpha_1$, while the duration after the negative slope is determined by $L_2^2/\alpha_2$[16]. Fig.5 also illustrates that in the case of a heat transfer structure with an infinitely thick second layer, the temperature curve exhibits a return to approximately -0.5 after the occurrence of a negative peak. This behavior differs from the case when the second layer thickness is finite, which is consistent with the majority of practical applications, where the temperature curve approaches 0 for sufficiently large values of $t$.

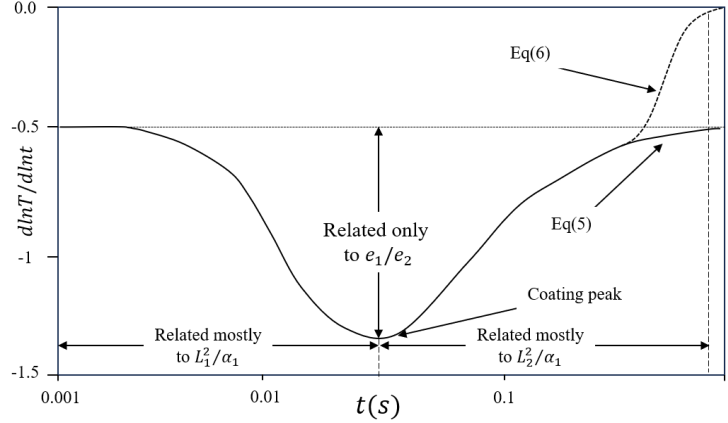

Fig. 5 Characteristics of logarithmic surface of temperature slope data for single-layer material

### 2.2.3 Thermal Tomography method

The Thermal Tomography (TT) method is frequently employed to analyze the thermal effusivity distribution of heterogeneous materials, allowing for characterization of material compositions and dimensions of various components within the 3D sample volume. The TT method involves correlating the measurement time with the depth within the sample, converting the measured surface temperature $T(x,y,t)$ at each surface position (x,y) into the material's thermal effusivity $e(x,y,z)$ as a function of 3D position $(x,y,z)$ inside the tested sample[22,23].

$$e(x,y,z) = \frac{d}{d\sqrt{t}}\left[\frac{Q}{T(x,y,t)\sqrt{\pi}}\right] \quad (7)$$

where $Q$ is the energy deposited by the flash on the surface position. The relationship between depth $z$ and time $t$ can be established as [23,24]:

$$z = \sqrt{\pi\alpha t} \quad (8)$$

where $\alpha(=k/\rho c)$ is thermal diffusivity, $\rho$ is density, and $c$ is specific heat, and $k$ is thermal conductivity. The TT method was found to satisfy the total effusivity conservation as well as to yield correct asymptotic values for single-layer and multilayer material systems [23]. The 3D thermal effusivity data obtained from the TT method can be further analyzed by slicing it into planes that are either parallel or perpendicular to the imaged surface. Additionally, a schematic diagram (Fig.5) is provided to illustrate the conversion of surface temperature data obtained from the bronze sample into 3D thermal effusivity data. It is important to note that the shape and magnitude of the predicted profiles are dominated by the material's thermal effusivity while the depth dimension is controlled by thermal diffusivity [23]. Therefore, it is necessary to know the thermal diffusivities of all the material layers when constructing 3D spatial thermal effusivity images from experimental data. Additionally, the parameter $Q$ is necessary for the determination of the absolute value of the material's thermal effusivity. It is related to surface heating intensity, the surface emissivity and surface roughness etc., which is difficult to be determined. However, since we are not interested in the absolute effusivity value but its depth distributions, $Q$ may be determined by assuming a constant thermal effusivity at the sample surface [23]. Similarly, due to the complexity of the corrosion composition of bronze ware, it is difficult to determine its thermal diffusivity. Here, the diffusivity of the rust layer is approximated as a constant at $\alpha_1$ =1.200 mm$^2$/s, and the diffusivity of the bronze substrate layer is set as $\alpha_2$ =9.775~12.45mm$^2$/s, which was measured from the sample made of materials with the same percentage composition. Nonetheless, it is still possible to differentiate between the cross-sectional features of a single-layer and a double-layer structure.

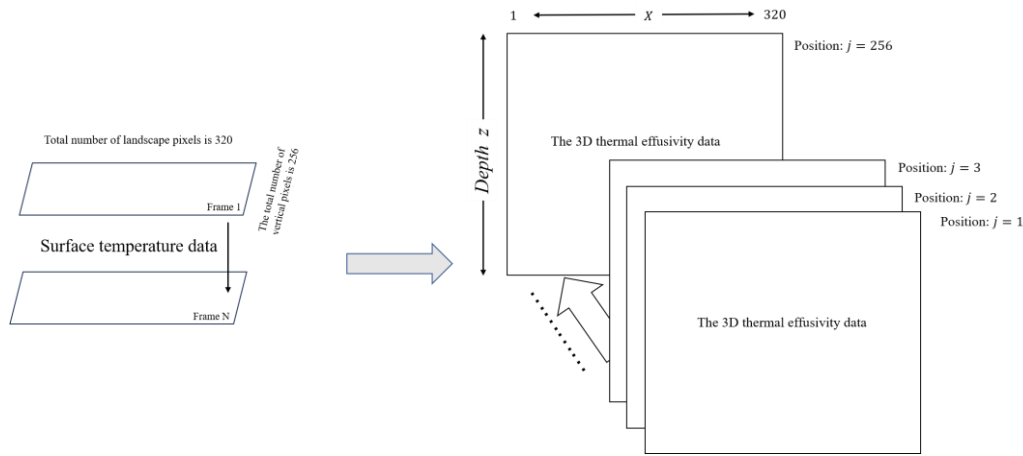

Fig. 6 A schematic diagram illustrates the conversion of the surface temperature data of a material into 3D thermal effusivity data.

In the experimental conditions outlined in Section 2.2, the resolution of the camera used was $x \times y = (320 \times 256)$. Consequently, the surface temperature data of the bronze specimen was divided into 256 groups based on the number of vertical pixels, with each group containing 3D thermal effusivity data ranging from position $x = 1$ to $x = 320$, based on the number of landscape pixels of the detector. Using the TT method, it is possible to accurately depict the variation in thermal effusivity at each point of the bronze surface with depth $z$, which makes it easier to differentiate the thickness of the rust layer from the metal layer.

## 3 RESULTS AND DISCUSSION.

### 3.1. The detection of the repaired patches

The infrared non-destructive testing image shown in Figure 6 reveals several thermal anomaly areas located on the left side of backside of the bronze ware cover. These thermal anomalies may be indicative of patching or repair materials added to the bronze after excavation. In Figure 6, we can observe 5 obvious thermal anomaly areas, numbered 1-5, which have been marked with red arrows in the figure. By comparing the visible light image on the right side of Figure 6, it is indicated that the green arrow marks a surface crack. The TT method was used to further analyze these anomalous structures and their thermal parameters. The thermal image results have i=320 pixels in the horizontal direction and j=256 pixels in the vertical direction. Place the graph in a coordinate system with horizontal coordinate $i$ and vertical coordinate $j$, $(i,j)$, so that each point in the result can be represented by coordinates.

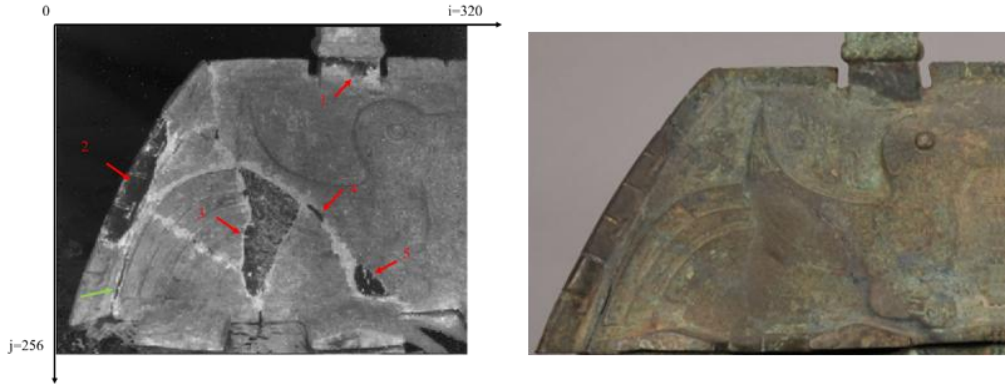

Fig. 7 Shows five obvious thermal anomaly areas and one obvious surface crack. The image on the right displays the visual appearance of this particular section of the artifact

For No.1 thermal anomaly we selected three sets of data ($j = 25, 28, 38$) and obtained the corresponding cross-sectional effusivity image using the TT method. In the effusivity image, the value of $z$ represents the relative depth of the point, and $z = 0$ means it is located on the surface of the bronze ware. $z = 100$ is the maximum relative depth present in this result. And there is a positive correlation between the grayscale value and the e-value in the results. the e-value of the bronze metal part in bronze vessels is greater than the e-value of the rust layer. In the results, along the z-axis, one can observe the grayscale variation of the e, which signifies the distribution of components at different relative depths. For bronze artifacts, their surfaces are covered with a layer of corrosion, whereas artificial repair materials do not possess this corrosion layer. Therefore, it is possible to differentiate between areas belonging to artificial repair materials and those inherent to the bronze artifact itself by examining the grayscale variations of e. For points (167,25), (235,25), and (170,28), situated at different locations on the bronze artifact, their curves all conform to the theoretical curve of a two-layer structure. Moreover, the negative peaks in the curves for points (167,25) and (235,25) occur at a similar time, indicating a comparable thickness of corrosion layer at these two points, which is consistent with their effusivity image Conversely, for point (170,28), the negative peak in its curve appears later than the previous two, aligning with the results from the effusivity image. For points (215,25) and (234,28), their curves conform to the theoretical curve of a single-layer structure, indicating that both points are located on a single-layer structure. This conclusion is further supported by the effusivity image. The difference lies in the effusivity image, where near point (215,25), one can still observe the corrosion layer alongside the metallic layer. In contrast, this observation is not present near point (234,28). The cause for this disparity is attributed to the irregular shape of the repair material. This viewpoint can be further validated through the thermal image. The red dashed line marked by j=38 passes below the thermal anomaly region in Fig.8c. As a result, a typical single-layer structure e distribution is not observable in the corresponding effusivity image. Instead, along the i-direction, a segment of continuous two-layer structure distribution is evident. In the effusivity image, point (126,38) exhibits the thickest corrosion layer, hence its curve displays a negative peak positioned further to the right along the time axis. Conversely, point (286,38) has the thinnest corrosion layer, resulting in its curve's negative peak being located furthest to the left. Point (143,38) features a corrosion layer thickness between the aforementioned two points, leading to its curve's negative peak falling between the peaks of the other two.

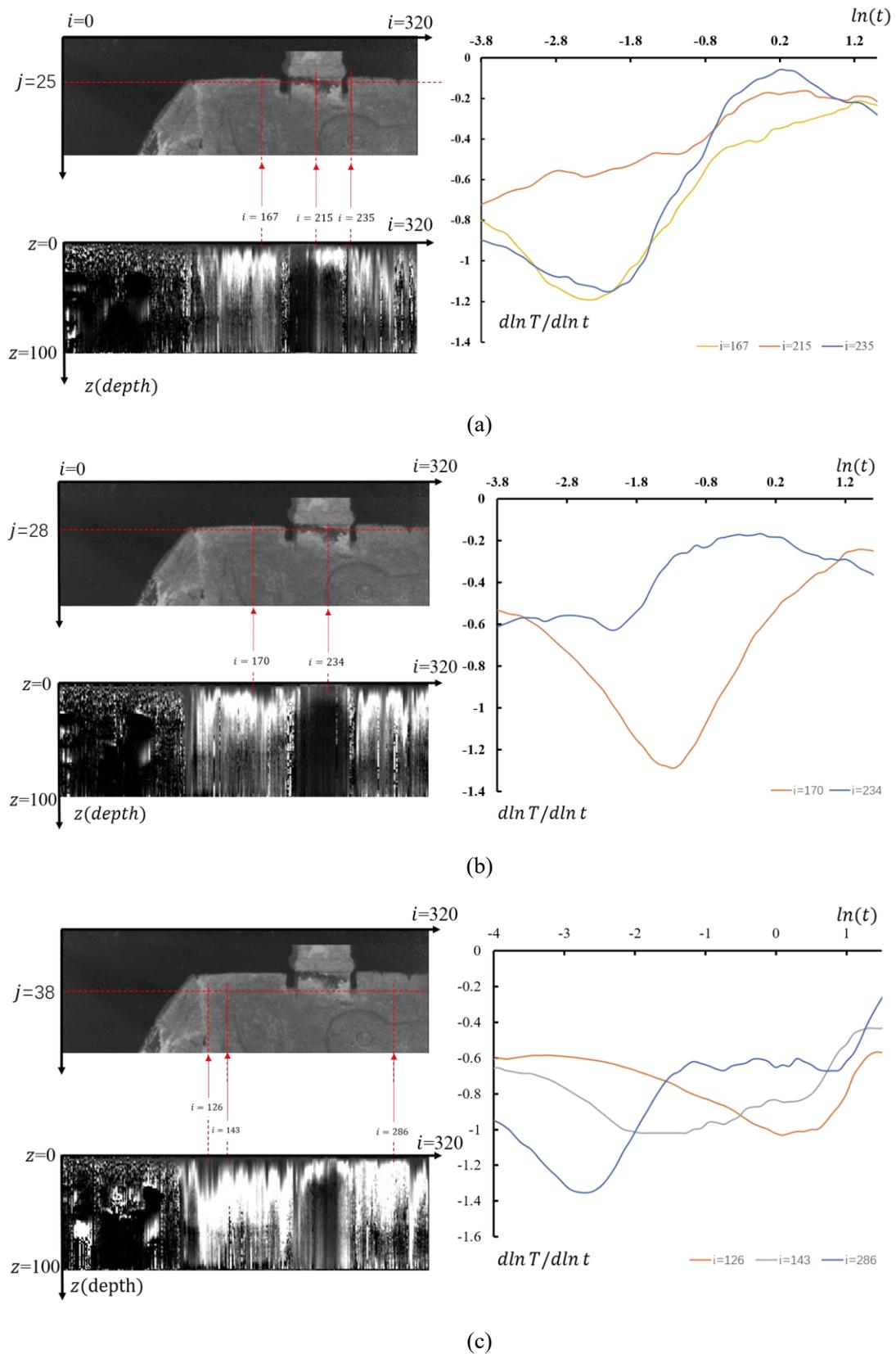

Fig. 8(a)(b)(c) depict three datasets ($j = 25, 28, 38$) of the No.1 thermal anomaly region.

We employed a similar method to investigate other regions of thermal anomalies, including the irregular shape marked in red in Figure 6, which is suspected to be caused by other repair materials, and the thermal anomaly area caused by surface cracks of the bronze ware in Figure 6. To investigate the characteristics of crack defects and repair materials, we analyzed the cross-sectional effusivity image of three red dotted lines, namely j=145,161,201, and the first-order logarithmic curve of some points on these dotted lines. The thermal and cross-sectional effusivity images of the bronze surface located at the red dotted line j=145 are presented in Figure 8. The dotted line j=145 intersects with three thermal anomaly areas, and the grayscale values are relatively low in the thermal image results. In Figure 8, we plotted the curve of the first-order logarithmic differentiation of some points on the dotted line j=145. The cross-sectional results of points (49, 145), points (177, 145), and points (202, 145) on the dotted line indicate a single-layer structure model. This observation suggests that the thermal properties of the repair materials at these points are isotropic.

The first-order logarithmic differentiation curves of the three points at the red dotted line j=145 in Figure 9 support the aforementioned observations. Notably, points (156, 145), (170, 145), and (177, 145) all belong to different surface positions of the same patch material. The cross-sectional results at points (156, 145) and (170, 145) reveal that the distribution of effusivity values at these positions is different from that at point (177, 145), as the effusivity gray levels at (156, 145) and (170, 145) are higher and close to that of the bronze metal layer. Therefore, we can conclude that the effusivity distribution of this triangular patch material appears uneven, unlike other repair materials used on the bronze. The distribution of effusivity in other parts of the repair material is generally uniform. Hence, it is likely that different repair materials and processes were employed in this part compared to other repair parts.

The point (73, 145) is of particular interest, as it is located at the intersection of the arc crack and the straight crack. By analyzing the cross-sectional image at this point, it was observed that the value of e near the surface is similar to that of the rust layer, whereas the distribution of e at a larger depth is similar to that of the metal layer. This suggests that the crack belongs to a two-layer heat conduction structure.

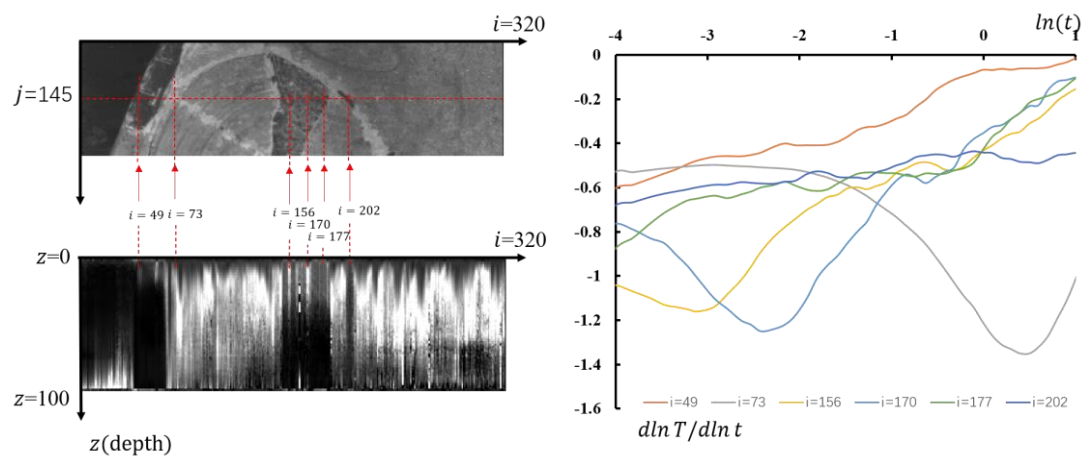

Fig. 9 Displays the cross-sectional effusivity image and thermal image of the bronze surface at the red dotted line j=145, and presents the first-order logarithmic differentiation curve of selected points on the same dotted line j=145.

Figure 10, the cross-sectional image of point (96, 161) shows that the crack at this site covers the rust layer, which is similar to the crack structure at point (73, 145). The cross-sectional images of points (96, 161) and (274, 161) also show cracks that cover the rust layer. From the TT results, it is evident that the thickness of the corrosion layer at point (96, 161) is smaller than that at point (274, 161). Furthermore, in the curve graph, the negative peak at point (96, 161) occurs earlier than at point (274, 161).Although points (154, 161) and point (156, 145) belong to the same repair part in the thermal image result, the distribution of e at point (154, 161) shows a single-layer heat conduction structure, indicating that the repair materials may also have certain areas with a two-layer structure, which is likely due to the repair techniques employed.

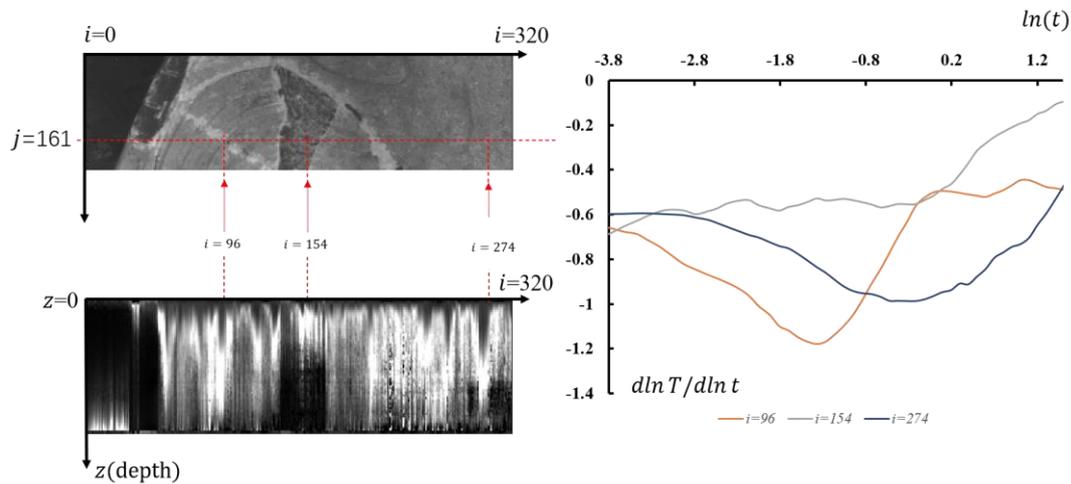

Fig. 10 The cross-sectional effusivity image and the thermal image of the bronze surface located at the red dotted line j=161, and the curve of the first-order logarithmic differentiation of some points on the dotted line j=161.

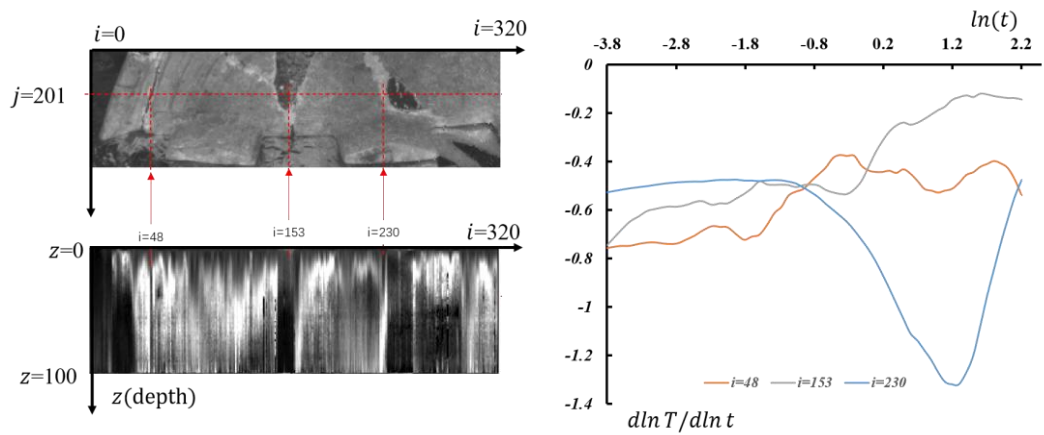

Fig. 11 the cross-sectional effusivity image and the thermal image of the bronze surface located at the red dotted line j=201, and the curve of the first-order logarithmic differentiation of some points on the dotted line j=201.

Figures 11, the thermal image result shows that the black thin strip at point (48, 201) does not exhibit significant peak characteristics in the corresponding curve. Upon closer examination, it is confirmed that this area corresponds to small cracks. The curve feature at point (153, 201) aligns with the characteristics of a single-layer structure, indicating it as a repair material region. At point (230, 201), significant peak features are observed in the curve, and the TT result also indicates a two-layer structure. Considering its proximity to the repair material region, it can be inferred that a filler was added at this location during the artifact restoration process, forming a two-layer structure over the original bronze substrate.

### 3.2. Surface corrosion analysis

The majority of the bronze artifact is enveloped by a loosely structured corrosion layer. However, in select areas, a fortuitous formation of a dense and durable oxide film has occurred, effectively preventing further corrosion of the internal bronze. Figure 11 showcases a portion of the bronze artifact where the oxide layer exhibits discontinuities, identifiable by lower grayscale values in the corresponding image. Typically, the corrosion of bronze artifacts occurs due to chemical reactions with water and carbon dioxide present in the surrounding air. These reactions lead to the formation of copper-containing salts, which are responsible for the majority of the observed corrosion on the bronze surface. However, in some fortuitous areas, a dense and compact layer of copper oxide forms, effectively acting as a protective barrier. This copper oxide layer inhibits further oxidation of the underlying bronze material by preventing the ingress of atmospheric oxygen.

Points 1 and 3 are situated on the surface exhibiting a black copper oxide layer, while Point 2 is located in a different region. Analysis of the corresponding curve graph reveals notable peak characteristics at Point 2, with the peak occurring towards the later stage, indicating a significant thickness of the corrosion layer compared to the bronze substrate. In contrast, no prominent peak features are observed at Points 1 and 3, suggesting that the corrosion layer in these areas can be considered negligible relative to the bronze substrate. Thus, the surface with the presence of a copper oxide layer can be regarded as a distinct single-layer structure.

Overall, it can be concluded that the corrosion process on the bronze artifact is predominantly driven by the formation of copper-containing salts, while the fortuitous presence of a dense copper oxide layer provides localized protection against further oxidation.

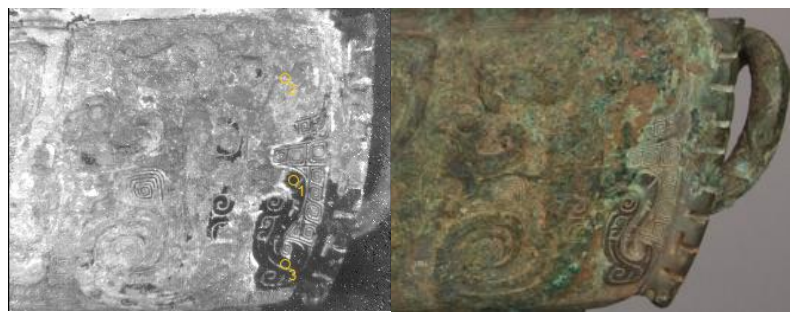

(a)

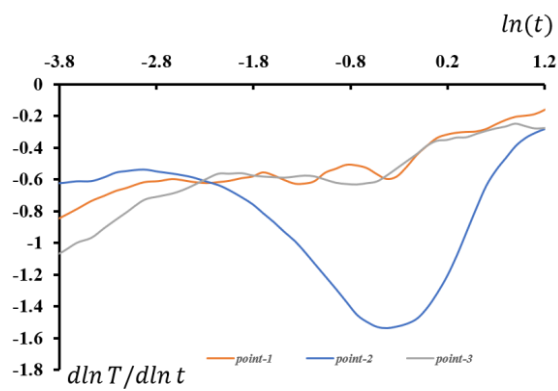

(b)

Fig. 12 The thermal image result with 3 points locations and displays the visual appearance of this particular section of the artifact. (b) shows the curve of the first-order logarithmic differentiation of those points.

The presence of copper oxide introduces challenges in determining the extent of the repair areas. However, these challenges can be mitigated by examining pseudo-color images. The results shown in Figure 12 provide evidence to support the aforementioned perspective.

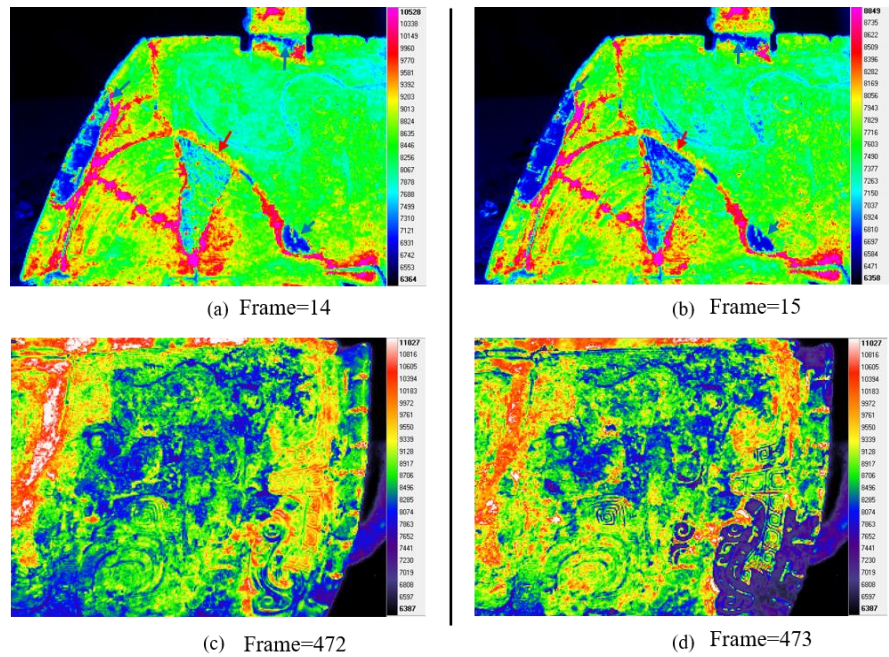

(a) Frame=14  (b) Frame=15
(c) Frame=472  (d) Frame=473

Fig. 13 Indicated the pseudo-colored raw results of the locations shown in Figures 6 and 13 Figures (a) and (c) are the pseudo-colored results of both images' flash frames, while Figures (b) and (d) are the pseudo-colored results of both images' frames immediately following the flash frames.

Through careful examination of Figures 13(a) and 13(c), notable observations can be made. Figure 13(a) reveals that during the occurrence of the flash, the temperature rise in areas containing repair materials is relatively lower compared to the surrounding regions of the bronze artifact. Conversely, in Figure 13(c), regions with a thin layer of copper oxide exhibit a temperature distribution that is relatively consistent with the adjacent areas.

After one frame of cooling, Figure 13(b) demonstrates that the repair area with a higher temperature (indicated by the red arrow) tends to approach the temperature of the surrounding repair area (indicated by the blue arrow). However, despite this convergence, a significant temperature difference between the repair area and the main body of the bronze artifact persists.

In Figure 13(d), the contour of the deep blue region (close to purple) shows a resemblance to the black contour observed in the grayscale image of Figure 12(a), while it is hardly visible in Figure 13(c). By comparing Figures 13(a)/(b) and 13(c)/(d), it becomes evident that the cooling rates differ noticeably between the repair materials and the copper oxide layer. Another obvious reason is that the obvious temperature difference around the repair material indicates that there is a crack structure around it (as indicated by the red arrow in Figure 13(a), the red temperature region around the repair material is irregular), while the oxide area around Figure 11(a) also has a red temperature area, but shows a clear pattern structure, corresponding to the surface pattern on the visible light picture. This indicates that no rupture has occurred in the surrounding area. It is feasible to employ similar methodologies to effectively distinguish between these components.

### 3.3. Surface cracks

The distribution of cracks on bronze artifacts carries valuable information for understanding their condition and historical context. Through the analysis of crack shapes and lengths, researchers can glean insights into the influences of environmental and external factors on the bronze object. Figures 14 and 15 present the results of the crack distribution on the surface of the bronze artifact's A and B sides, respectively. Due to the limited field of view of the thermal imager, the results presented in Figure 15 and Figure 15 are obtained by stitching together multiple partial results. All the results have similar or identical frame numbers, ensuring that they represent the surface of the bronze artifact at the same or similar depth.

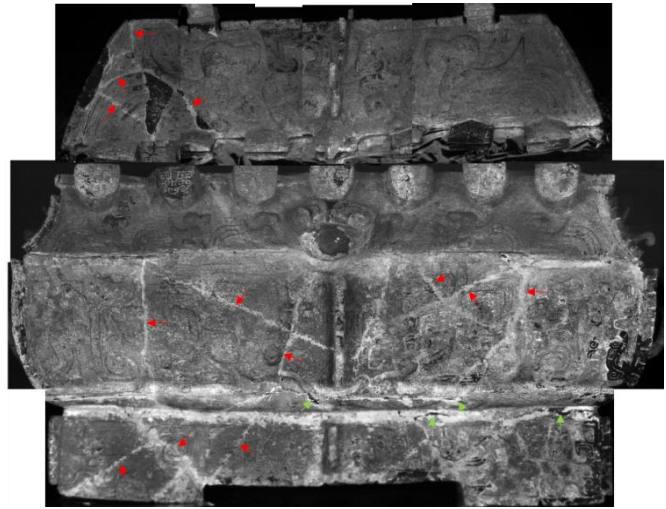

Fig. 14 Distribution of cracks on the backside surface of the bronze artifact.

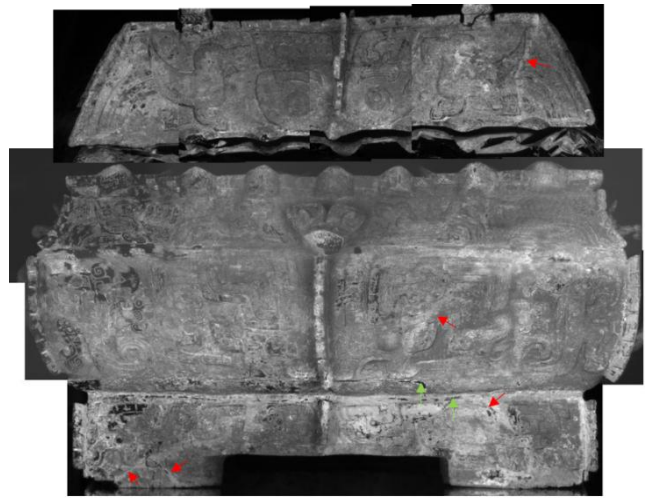

Fig. 15 Distribution of cracks on the frontage surface of the bronze artifact.

Figure 14 and 15 displays the distribution of cracks observed on the backside and frontage of the bronze artifact. The image provides a visual representation of the spatial arrangement and characteristics of the cracks. The cracks on the bronze artifact typically exhibit regular straight or curved shapes and appear in a grayish-white color. These characteristics are indicated by red arrows in Figure 13 and Figure 14. For the backside surface, the cracks are primarily concentrated on the left side of the lid and surround the patch area. In the body of the artifact, several distinct grayish-white straight and curved lines can be observed, indicating that this side of the bronze artifact is composed of several large fragments that have been welded, repaired, and joined together. As for the frontage surface, the distribution of cracks is less prominent than on the backside surface. These cracks are mainly located on the right side of the lid, forming an irregular zigzag line that traverses the entire lid from top to bottom. There is also a less noticeable crack distribution in a northeast direction on the body of the artifact. However, this crack does not extend vertically across the entire body. The cracks in the foot section are not as apparent, but based on the distribution of grayish-white areas, it can still be confirmed that there are two cracks present (indicated by red arrows in Figure 14 at the foot of the artifact). The green arrows in Figure 13 and Figure 14 indicate some gaps that have formed during the joining process of the bronze artifact. It is worth noting that there are many such gaps in the artifact, and the green arrows represent only a portion of them.

## 4 CONCLUSIONS

This study applied pulsed thermal imaging (PT) combined with thermal tomography (TT) for the non-destructive examination and evaluation of the late Shang dynasty "Fu Hao" bronze artifact. Based on the theoretical framework of single-layer and double-layer heat conduction models, the approach enabled spatial localization of restoration interventions such as repair patches and filler materials, while effectively differentiating restorative substances from the original bronze substrate. The main findings are summarized as follows.

Thermal tomography proved capable of identifying major repair areas through depth-resolved effusivity profiles. Most repaired regions exhibited a single-layer thermal structure, in contrast to the typical "corrosion layer–metal substrate" double-layer configuration of the original bronze, indicating a distinct thermal property mismatch between the non-metallic restoration materials and the metallic base. One repair zone showed an effusivity distribution in the depth direction similar to that of the metal layer, suggesting the possible use of metal-containing material or a different restoration technique in that area.

The surface corrosion of the artifact was non-uniform. While most of the surface was covered with a loose corrosion layer, localized areas had formed a dense oxide film (e.g., copper oxide). These oxide-coated regions behaved thermally as a near-single-layer structure and displayed cooling characteristics distinct from the surrounding corroded areas, implying that the compact oxide layer inhibited further corrosion progression.

Surface cracks were resolved as double-layer structures comprising the corrosion layer and the metal substrate, confirming that the cracks penetrated through the surface patina into the underlying bronze. The distribution and morphology of the cracks reflected the stress history and preservation conditions of the artifact, with a higher density of cracks on the back side consistent with its burial orientation and restoration history.

The integration of pulsed thermal imaging with thermal tomography provided three-dimensional visualization of thermal properties beneath the surface without physical contact. The method is sensitive to non-metallic repair materials, thin oxide films, and shallow cracks, making it suitable for diagnosing complex, layered structures in cultural heritage objects.

The main limitation of the approach lies in the reduced accuracy in retrieving the thickness of the metal substrate at greater depths due to thermal diffusion effects. Spatial variation in the thermal properties of the corrosion layer also introduces uncertainty. Future work should combine multi-modal non-destructive data (e.g., X-ray imaging, multispectral imaging) and incorporate measured thermophysical parameters of typical corrosion and restoration materials to improve quantitative reliability.

In summary, pulsed thermal imaging coupled with thermal tomography serves as an effective diagnostic tool for bronze cultural heritage, delivering critical information on restoration traces, corrosion distribution, and interfacial structures that can inform conservation strategies and historical interpretation.